# High Resolution Infrared Imaging of FSC 10214+4724: Evidence for Gravitational Lensing[1]

James R. Graham and Michael C. Liu

Astronomy Department, University of California Berkeley, CA 94720-3411

jrg/mliu @ astro.berkeley.edu

astro-ph/9506030   5 Jun 1995

---

## ABSTRACT

We present near–infrared observations of the ultraluminous high–redshift ($z = 2.286$) IRAS source FSC 10214+4724 obtained in 0."4 seeing at the W. M. Keck telescope. These observations show that FSC 10214+4724 consists of a highly symmetric circular arc centered on a second weaker source. The arc has an angular extent of about 140° and is probably unresolved in the transverse direction. This morphology constitutes compelling *prima facie* evidence for a gravitationally lensed system. Our images also contain evidence for the faint counter image predicted by the lens hypothesis. The morphology of FSC 10214+4724 can be explained in terms of a gravitationally lensed background source if the object at the center of curvature of the arc is an $L^*$ galaxy at $z \approx 0.4$.

If FSC 10214+4724 is lensed then there is significant magnification and its luminosity has been overestimated by a large factor. Our results suggest FSC 10214+4724 is not a uniquely luminous object but ranks among the most powerful quasars and ultraluminous IRAS galaxies.

*Subject headings:* Galaxies: individual (FSC 10214+4724) – galaxies: starburst – infrared: galaxies

## 1. Introduction

Since its serendipitous discovery by Rowan–Robinson *et al.* (1991), the IRAS source FSC 10214+4727 (hereafter F 10214) at a redshift of 2.286 has earned a position as one of the most tantalizing objects in the high redshift universe. With an inferred rest frame luminosity of $\approx 3 \times 10^{14} L_\odot$, it is the most luminous IRAS galaxy and ranks among the most luminous objects known in the Universe. It has been suggested that F 10214 is a protogalaxy, a massive gas cloud undergoing dissipative collapse and a global primordial starburst (*e.g.*, Rowan-Robinson *et al.* 1991, Solomon, Radford, & Downes 1992; Solomon, Downes, & Radford 1992). Enormous continuum emission from far-IR to mm wavelengths indicate substantial quantities of dust heated by an gigantic starburst (Rowan-Robinson *et al.* 1991, Soifer *et al.* 1991, Downes *et al.* 1992, Clements *et al.* 1992, Soifer *et al.* 1992, Rowan-Robinson *et al.* 1993, Matthews *et al.* 1994). Evidence for vigorous star formation is supported by the detection of high excitation CO line emission, suggesting the presence of substantial quantities of interstellar gas (Brown & Vanden Bout 1992, Solomon *et al.* 1992ab, Kawabe *et al.* 1992). Rest frame UV and optical spectra reveal a wide variety of excitation conditions without any evidence for very broad line emission (Rowan-Robinson *et al.* 1991, Elston *et al.* 1994, Soifer *et al.* 1995). Combined with its unusually high rest frame UV polarization (Lawrence *et al.* 1993, Januzzi *et al.* 1994), the data suggest FSC 10214 harbors an obscured active nucleus similar to Seyfert 2 nuclei in the local universe (Elston *et al.* 1994).

Previous IR imaging with the Keck telescope in $0.''6 - 0.''8$ seeing shows that F 10214 has an unusual morphology, consisting of a bright C–shaped structure near a complex of fainter sources (Matthews *et al.* 1994). Matthews *et al.* attribute this morphology to a tidal interaction. An alternate interpretation is that the source $\simeq 1''$ N of F 10214 is a foreground galaxy which acts as a gravitational lens (Matthews *et al.* 1994, Elston *et al.*

1994, Trentham 1994, Broadhurst 1995). Elston *et al.* (1994) note that the companions of F 10214 have colors and magnitudes characteristic of foreground galaxies rather than objects at $z = 2.3$. Gravitational lensing by one of these companions would provide a natural explanation of the arc–like morphology and could explain the elongated and extended nature of the emission line region. The difference in morphology between 2.12 and 2.16 $\mu m$ led Matthews *et al.* to argue against the lensing hypothesis. As arcs are only formed when an extended source is lensed, the difference between the continuum and the H$\alpha$ line may simply reflect structure in the source.

In this *Letter* we present new IR data for F 10214 with substantially improved spatial resolution. Our observations favor the gravitational lens model. Throughout this discussion we will use $\Omega_0 = 0$ and $H_0 = 75 \ km \ s^{-1} \ Mpc^{-1}$. With this choice of cosmology, $1''$ corresponds to 8.8 kpc.

## 2. Observations and Reductions

We observed F 10214 on 1995, March 18 (UT) using the 10–m W. M. Keck telescope, Mauna Kea, Hawaii, with the facility near–IR camera (Matthews & Soifer 1994). The camera is equipped with a Santa Barbara Research Corporation 256 × 256 InSb array. The pixel size is $0.''15$. We observed with the K (2.0 – 2.4 $\mu m$) filter. An integration time of 10 $s$ per frame was used. Six frames were coadded before the telescope was offset by a few arcseconds. The telescope was stepped using a non-redundant dither pattern. Each exposure was guided using an off–axis CCD camera.

Reduction of each coadded frame consisted of subtracting a sky frame constructed by averaging prior and subsequent frames. Objects in the sky frames were identified and excluded from the average. Sky–subtracted frames were then flat–fielded using an average of dark–subtracted twilight sky frames. We measured the positions of two to three stars common to all the reduced frames to determine the spatial registration and then shifted individual frames by integer pixel offsets to assemble a mosaic of the field. A total of thirty frames (1800 $s$) were obtained, but the seeing degraded as the observation progressed. We selected the twenty best frames, chosen on the basis of the full width at half maximum (FWHM) of the bright star to the south of FSC 10214+4724 (star A of Rowan–Robinson *et al.* 1991). In the individual frames the FWHM ranges from 0."41 to 0."5. The final mosaic is shown in Figure 1. The mosaic achieves a 1 $\sigma$ limit of 23.9 magnitudes per square arcsecond and has a total integration time of 1200 s. The images in the mosaic are seeing limited with 0."44 FWHM.

## 3.   Morphology of F 10214

Figure 1 shows that F 10214 consists of several components. There is a bright extended arc to the south (source 1, following Matthews *et al.* 1994). Redshifted H$\alpha$ falls within the K filter and makes a 40% contribution to the observed flux in a 0."8 box centered on the arc; other weaker emission lines contribute another 10% (Soifer *et al.* 1995). The arc is the sole source of the optical line emission and is associated with the radio source which is also elongated EW (Soifer *et al.* 1992, Lawrence *et al.* 1993, Soifer *et al.* 1995). The arc defined by source 1 is centered on a fainter source (source 2). There are two faint companions to the north-northeast (source 3 and 4; the former is closer to sources 1 and 2). The arc is marginally resolved with a thickness $\lesssim$ 0."3. Source 2 has a size of at least 0."5, but it is hard to define its size because it sits on a plateau of faint emission which extends over a radius of about 1". This weak extended emission is centered on source 2 and is probably associated with it. Source 2 also shows a component which extends off to the north. Matthews *et al.* (1994) note that source 3 is extended in the east-west direction; we clearly resolve it into two components. Source 4 is also extended.

As the signal to noise in these data is high (about 60 at the peak of source 1) we chose to sharpen the image by using the maximum entropy deconvolution as implemented in the STSDAS IRAF package. We also used Lucy-Richardson deconvolution to check the uniqueness of the deconvolution; both procedures resulted in nearly identical reconstructions. The deconvolved image of F 10214 is presented in Figure 2 which shows the morphology described above with increased clarity. The most striking aspect of Figure 2 is the incomplete circular arc of 140° extent with an angular radius of 1."2 (11 kpc) centered within 0."1 of source 2. The arc is asymmetric: the brightest point is not at the mid–point of the arc but is off to the east.

F 10214 is unlike any normal galaxy. Matthews *et al.* (1994) point out that it resembles the luminous ($3 \times 10^{11} L_\odot$) interacting system Arp 148. The similarity between F 10214 and Arp 148 is superficial in the light of this new data. The Arp 148 plate (Arp 1966) shows that it is a ring galaxy (Burbidge 1964). The features that Matthews *et al.* associate with the arc of F 10214 are not tidal tails, but part of a ring. Pursuing this analogy further, sources 2 and 3 are the nuclei of the original interacting galaxies. The unsatisfactory aspect of this picture is that the ring, and not the deep nuclear potential well, is the seat of the activity in F 10214. This prejudice is confirmed by IR and radio observations of Arp 148, which locate the IR source and a compact radio core ($\theta = 0."2 \times 0."1$) in the optical nucleus (Joy & Harvey 1987, Condon *et al.* 1991). We therefore find that it is unlikely that the morphology of F 10214 can be explained in terms of tidal features.

The morphology and symmetry of F10214 is reminiscent of gravitationally lensed objects, especially the rings produced by lensing of high redshift radio lobes (Hewitt *et al.* 1988; Langston *et al.* 1989; Jauncey *et al.* 1991; Patnaik *et al.* 1993) and the large scale optical arcs produced by gravitational lensing in clusters (Luppino *et al.* 1993, Melnick *et al.* 1993, Fort 1992). In the next section we show that it is plausible that F 10214 is gravitationally lensed. Ultimately, redshifts for the individual components will establish whether F 10214 is an interacting system or lensed.

## 4. FSC 10214+4724 as a Gravitational Lens

A unique lens model for F 10214 cannot be constructed because of the limited data available (Trentham 1994). However, we can show that the observed morphology is consistent with lensing. The model presented here is based on the suggestion (Broadhurst 1995) that the northen extension of source 2 is a second image of F 10214. Broadhurst & Lehar (1995) have proposed a detailed model which is similar to the sketch presented here.

The imaging properties of gravitational lenses are discussed in Blandford and Narayan (1986, 1992). If the lens has circular symmetry and an extended source lies exactly along the optical axis the resultant image is an Einstein ring. If the source is displaced from the symmetry axis, then two arcs of equal length will be formed. Failure to observe two equal arcs, as in the current case, implicates a more realistic asymmetric lens (Grossman & Narayan 1988).

The source plane of a gravitationally lensed object can be divided into regions based on the image multiplicity; the curves separating these regions are caustics. Sources which lie near caustics produce highly magnified images. The conjugates to the caustics in the image plane are known as critical lines. The morphology and brightness of the arc in F 10214 suggest that source 2 is the lensing galaxy and that the background object is located close to a cusp of the tangential caustic, which separates regions of image multiplicity 3 and 5 (see Figures 6 and 7 of Blandford and Narayan 1992). Given this assumed lens configuration, the resultant image for a background point source consists of three highly magnified images distributed in an arc along the tangential critical curve. A fourth image is highly demagnified and coincides with the lens; a fifth image should lie along the line of mirror symmetry defined by the arc and lensing galaxy. The fifth image should be slightly displaced to the north of the lensing galaxy (source 2). See Figure 3 for a sketch of the proposed lens configuration. There is clearly a northern extension to source 2 (§3). We removed the lensing galaxy by subtracting a scaled point spread function centered on source 2. The result is shown in Figure 4. A faint blob is seen at a position angle of about $-10°$, roughly 0."3 from the center of source 2. This is probably the fifth image produced by the slightly off–cusp source.

If the background source is extended, the images can be distorted into highly elongated arcs when the source lies on a caustic (Narayan & Wallington 1992). This effect is seen in lensed radio lobes or in clusters with gravitational arclets. Given the smoothness and large extent of the arc of F 10214 and the difference in appearance between the continuum (2.12 $\mu m$) and H$\alpha$ (2.16 $\mu m$), the background source must be extended. The asymmetry of the arc of F 10214 (§ 3) also suggests that the source galaxy is located slightly off the cusp of the tangential caustic (see Figure 2 of Wallington and Narayan 1993).

A simple quantitative argument demonstrates the viability of the lens interpretation. Since the redshift of the lens is unknown, we will assume that the lens is located at the most likely distance for a lensing event to occur. Assuming Euclidean geometry and a constant comoving density of potential lenses, the probability for lensing is maximized when $D_l = D_s/2$, where $D_l$ and $D_s$ are the angular size distances for the lens and source, respectively, as seen by the observer (Gott *et al.* 1989). If we model the lens as a singular isothermal sphere with one-dimensional velocity dispersion $\sigma$ at $D_l = D_s/2$ then the angular separation, $\theta$, of the lensed images is

$$\theta = 4\pi \left(\frac{\sigma}{c}\right)^2 \approx 0."289 \left(\frac{\sigma}{100 \ km \ s^{-1}}\right)^2,$$

(Blandford and Narayan 1992). Taking $\theta \approx 1."5$ from the observed separation of the arc and counter image implies $\sigma \approx 230 \ km \ s^{-1}$. This velocity dispersion corresponds approximately to an $L^*$ galaxy (Faber & Jackson 1976, Aaronson *et al.* 1986). Assuming $M_B^* = -19.7$, an $L^*$ galaxy at $D_l = D_s/2$ (*i.e.* $z = 0.35$) has $K = 16.7 - 17.3$ depending on whether it is an elliptical or spiral. (K–corrections were calculated using the spectral energy distributions of Bruzual and Charlot [1993].) Photometry of the putative lens (source 2) from our images gives $K = 17.6$ in $1.''8$ diameter beam. Applying an aperture correction from Aaronson (1977), based on assuming $D_0 = 45\ kpc$, gives a total $K$ magnitude of 16.9. This lies well within the range expected for the lensing galaxy.

## 5. Implications

If F 10214 is gravitationally lensed, its total luminosity has been overestimated. The length of the arc (30 $kpc$) is consistent with lensing of a kpc sized region (as implied by the transverse thickness of the arc) if the magnification factor is $\approx 10$. Trentham (1994) shows that the probability that the magnification is between two and ten is about 25%. Broadhurst & Lehar (1995) conclude that the magnification is $\simeq 50$, and that F 10214 is similar to local Seyfert II galaxies such as NGC 1068. However, very large magnifications (> 20) are unlikely (Trentham 1994). Assuming that the observed flux is amplified by an order of magnitude, the bolometric luminosity for F 10214 drops to about $10^{13}\ L_\odot$, in line with the rest of the ultraluminous IRAS galaxy sample (Sanders *et al.* 1988a and 1988b). This makes it unnecessary to invoke an exotic explanation, such as a primordial global starburst, to account for the luminosity of F 10214. The inferred mass of molecular gas is reduced to $10^{10} M_\odot$, comparable to quantities measured in other ultraluminous IRAS galaxies; there is no longer any reason to believe that F 10214 is a galaxy composed primarily of molecular gas (Blitz 1992, Clements *et al.* 1992, Downes *et al.* 1992, Solomon *et al.* 1992ab). Since virial mass estimates depend on the size of the emitting region, the mass inferred from the CO linewidth of $240\ km\ s^{-1}$ (Solomon *et al.* 1992ab) must also be reduced proportionately to about $10^{10}\ M_\odot$.

The amplification due to gravitational lensing may be frequency dependent if the source is extended (§4); regions which dominate the emission at different wavelengths will be lensed differently according to their spatial distribution. However, we do not expect there to be gross differences in the amplification of the IR and sub-millimeter continuum and millimeter line emitting regions since these are likely to be cospatial. Thus the overall shape of the intrinsic spectral energy distribution at mid-IR and longer wavelengths should not be much different than that observed. The spectral energy distribution is similar to galaxies in the warm ultraluminous IRAS sample ($f_\nu(60~\mu m)/f_\nu(100~\mu m) > 0.2$, Sanders *et al.* 1998b), especially to the warmest objects in this category such as the radio quiet quasars PG 0050+124 (1 Zw 1) and PG 0157+0001 (Mrk 1014). The gas and dust masses are comparable to other ultraluminous IRAS galaxies (*cf.*, Table 2 in Solomon *et al.* 1992b). The mass–ratio of gas to warm dust, as derived from the CO line and sub-millimeter continuum, is comparable to other systems ranging in luminosity from M82 to Arp 220 (Downes *et al.* 1992). Therefore, our results strongly suggest that F 10214 is a typical ultraluminous IRAS galaxy that is lensed by a foreground galaxy.

Kormendy and Sanders (1992) suggest that elliptical galaxies are formed by the merger induced dissipative collapse that occurs when gas rich disk galaxies collide. In the local Universe these mergers are the ultraluminous IRAS galaxies (Sanders *et al.* 1988a). If F 10214 is a high redshift analog of the ultraluminous IRAS galaxies then it may be the first example of an giant elliptical galaxy forming at high redshift.

It is a pleasure to acknowledge Tom Broadhurst's key role in renewing our interest in the possibility that F 10214 is gravitationally lensed.

The W. M. Keck Observatory is a scientific partnership between the University of California and the California Institute of Technology, made possible by the generous gift of the W. M. Keck Foundation and support of its president, Howard Keck. It is also a pleasure to thank W. Harrison and J. Aycock for their help with these observations.

JRG is supported by a Packard Foundation fellowship, and MCL is supported by a NSF graduate student fellowship.

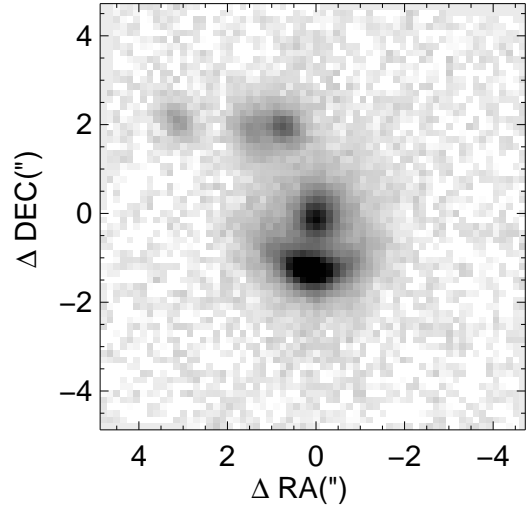

Fig. 1.— The Keck K–band images of FSC 10214+4724 displayed as a logarithmic grey scale.

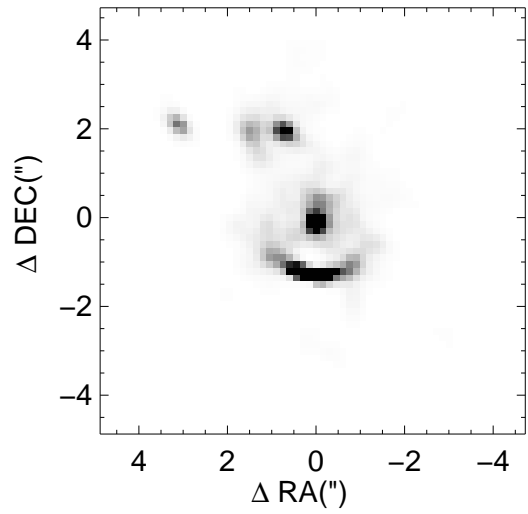

Fig. 2.— The deconvolved K–band image.

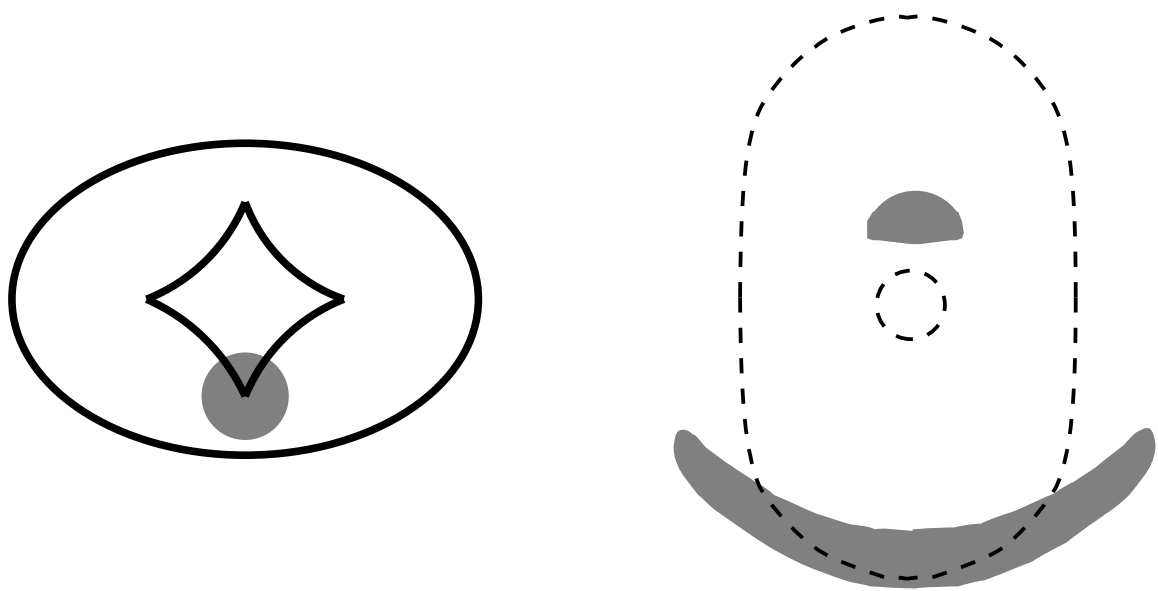

Fig. 3.— A sketch of the proposed elliptical lens geometry. The left hand figure shows the lens caustics of an elliptical lens in the source plane. The shaded circle represents the source. The right hand part of the figure represent the image plane. The dotted lines on the right are the critical curves, and the shaded areas are the magnified images of the source.

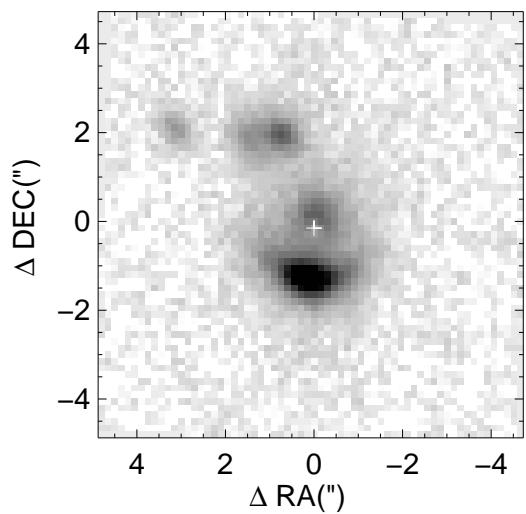

Fig. 4.— The nortern extension to source 2 is emphasized by subtraction of a point source centered on the cross – the location of source 2. A faint blob is seen to the N roughly 0."3 from the center of source 2. This is probably the fifth lensed image of FSC 10214+4724 (see §4).